\documentclass{icrc}
\usepackage{times}
\usepackage{graphicx} % when using Latex and dvips
\begin{document}

\title{Study of the TeV gamma-ray spectrum of SN 1006 around the NE Rim}
\author[1]{T. Tanimori}
\affil[1]{
Department of Physics, Graduate School of Science, Kyoto University,
Kyoto, 606-8502,  Japan
}
\author[2]{T. Naito}
\affil[2]{Faculty of Management Information, Yamanashi Gakuin University, Kofu, Yamanashi 400-8575, Japan}
\author[3]{T. Yoshida}
\affil[3]{Faculty of Science, Ibaraki University, Mito, Ibaraki 310-8512, Japan}
\author[4]{P.G. Edwards}
\affil[4]{Institute of Space and Astronautical Science, Sagamihara, Kanagawa 229-8510, Japan}
\author[5]{S. Gunji}
\affil[5]{Department of Physics, Yamagata University, Yamagata, Yamagata 990-8560, Japan}
\author[6]{S. Hara}
\affil[6]{Department of Physics, Tokyo Institute of Technology,
Meguro-ku, Tokyo 152-8551, Japan}
\author[2]{T. Hara}
\author[7]{A. Kawachi}
\affil[7]{Insitute for Cosmic Ray Research, University of Tokyo,
Kashiwa, 277-8582 Chiba, Japan}
\author[7,8]{T. Kifune}
\affil[8]{Faculty of Engineering, Shinshu University, Nagano, Nagano 380-8553, Japan}
\author[9]{Y. Matsubara}
\affil[9]{STE Laboratory, Nagoya University, Nagoya, Aichi 464-8601, Japan}
\author[10]{Y. Mizumoto}
\affil[10]{National Astronomical Observatory of Japan, Mitaka, Tokyo 181-8588, Japan}
\author[7]{M. Mori}
\author[6]{M. Moriya}
\author[11]{H. Muraishi}
\affil[11]{Ibaraki Prefectural University of Health Sciences,
 Ami, Ibaraki 300-0394, Japan}
\author[9]{Y. Muraki}
\author[12]{K. Nishijima}
\affil[12]{Department of Physics, Tokai University, Hiratsuka, Kanagawa 259-1292, Japan}
\author[13]{J.R. Patterson}
\affil[13]{Department of Physics and Math.\ Physics, University of
Adelaide, 
SA 5005, Australia}
\author[6]{K. Sakurazawa}
\author[7]{R. Suzuki}
\author[14]{T.Tamura}
\affil[14]{Department of Physics, Kanagawa University, Yokohama, Kanagawa 221-8686, Japan}
\author[3]{S. Yanagita}
\author[15]{T. Yoshikoshi}
\affil[16]{Department of Physics, Osaka City University, Osaka, Osaka 558-8585, Japan}

\correspondence{T.Tanimori (tanimori@cr.scphys.kyoto-u.ac.jp)}

\firstpage{1}
\pubyear{2001}

\titleheight{11cm} % uncomment and adjust in case your title block
                     % does not fit into the default and minimum 7.5 cm

\maketitle

\begin{abstract}
The differential spectrum of TeV gamma rays between 1.5 TeV and 20 TeV
from the north-east rim of SN1006 was obtained
from the data observed  in 1996 and 1997
using  the 3.8m CANGAROO \v Cerenkov telescope.
This spectrum  matches the model calculated using  the Inverse Compton (IC)
process with 2.7k Cosmic Microwave Background (CMB).
This enables  us to estimate the absolute strength of
the magnetic field around the shock and the maximum energy of
accelerated electrons with the considerable accuracy:
the obtained field strength and maximum electron energy are
$4\pm1$ $\mu$G and 50 TeV respectively.

Also we have detected again the TeV gamma-ray emission from the same
position using the 10m CANGAROO-II telescope in 2000,
and the preliminary spectrum around 1 TeV region is 
presented in this conference. 
The two spectra agree well in the overlapped energy region. 
\end{abstract}

\section{Introduction}

Recent observations in both X rays and TeV gamma rays
dramatically revealed the existence of non-thermal particles
in the Supernova Remnants (SNR),
which are accelerated up to $\sim$ 100 TeV presumably 
by the shock acceleration process.
First  set of such observations is the intense non-thermal X-ray emission
from the rims of Type Ia SNR SN1006 (G327.6 +14.6)
by ASCA (Koyama et al.\ 1995) and ROSAT (Willingale et al. 1996),
and the subsequent detection of  
TeV gamma-ray emission from the Norh East (NE) rim 
by CANGAROO (Tanimori et al. 1998).
The detected integral fluxes  above TeV were consistent with 
predictions  based on the Inverse Compton scattering (IC) of 
high energy electrons with low
energy photons of 2.7 K cosmic background
(Mastichiadis and de Jager 1996; Yoshida and Yanagita 1997).
Following this success, other two SNRs have been detected
in the same two energy bands;
strong synchrotron X-ray emission and TeV gamma rays 
from RXJ1713.7$-$3946 (G347.3$-$0.5) were 
detected by ASCA in 1997 (Koyama et al., 1997) and by CANGAROO in 2001
(Muraishi et al., 2001) respectively, and 
also those from Cassiopeia A were detected by Beppo SAX (Allen et al., 1997) 
and HEGRA (P\"uhlhofer et al., 1999). 

At present,  SNRs are generally favored as the site for the generation
of galactic cosmic rays (mainly protons).
High energy gamma rays detected from SNRs could be considered
to emanate from $\pi^{\circ}$  decays induced by collisions
between swept-up matter and accelerated protons in SNRs.
On the other hand, 
gamma-rays detected from SN1006 are likely to be explained
by IC radiation by very high energy electrons (Tanimori et al. 1998),
because the matter density around  the shock front in SN 1006
is too tenuous ($\sim$0.4 cm$^{-3}$: Willingale et al. 1996) to
generate the observed TeV gamma-ray flux from $\pi^{\circ}$s decays. 
The model based on the synchrotron-IC process
can  fit all the data of radio, X-ray and TeV gamma rays.
This model, in addition, predicts a flatter   
power law spectrum ($E^{s}dE$) 
with an index of $ s =\sim  -1.6$ below the TeV band with 
a turnover to a steeper spectra around the TeV region.
In this scheme, this steepening can be  explained
by a high energy limit of parent electrons,
as an energy flux peak of the synchrotron emission 
appears in the soft X-ray region.

At present there still remains the possibility
that an additional component of gamma rays from $\pi^{\circ}$
decays with the index of $s \sim -2.2 $
would gradually dominate instead of IC radiation in the sub-TeV energy 
region.
Detailed  study of the spectral shape is quite important to 
answer this question.

In April 2000, the new 10m CANGAROO-II telescope 
was completed (Tanimori, et al., 2001 and Kawachi et al., 2001).
Both RXJ1713.7$-$3946 and SN1006 have been detected again with high statistics
by this new telescope,  
and their results are presented in this conference 
(Enomoto et al, 2001; Hara et al., 2001).
Thus we have firmly established the TeV gamma-ray emission from SNRs.

Here we report on a further analysis of the same data set
used in the previous paper (Tanimori et al., 1998)
in order to extract spectral information for the reasons
mentioned above.

%The differential flux presented here was already obtained in 1999
%using the same data for Tanimori et al. (1998), but at that time 
%the negative result for the detection of TeV gamma rays of SN1006 
%was presented (Chadwick et al., 2000).
%Therefore we have keep it back until the reconfirmation of the
%emission of TeV gamma rays.
%Now we reveal this significant data.

\section{Analysis and Results}
Observations were done with the 3.8m 
\v Cerenkov imaging telescope of the CANGAROO
Collaboration (Patterson \& Kifune 1992; Hara et al. 1993)
near Woomera, South Australia 
(136$^{\circ}$47' E and 31$^{\circ}$06'S) in 1996 and 1997.

\begin{figure}[t]
\vspace*{2.0mm} 
\includegraphics[width=9.3cm]{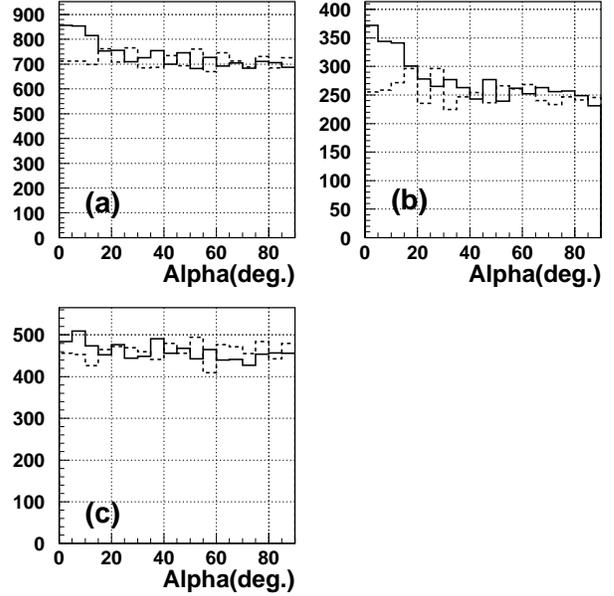} 
\caption{
(a) Plot of the event distribution as a function of the
$\alpha$  for the 1997 data.
(b) is the similar plot after applying the timing  selection, 
and (c) is also the similar plot for rejected events
of (a) by this cut.
}
\label{fig:1}
\end{figure}

An imaging analysis using the conventional  parameterization 
was applied for the data, where  the parameter cuts
were varied as a function of the energy of the shower.
These variations of parameter cuts were estimated from the simulation
study. 
Here we add the additional analysis to enhance the gamma-ray signals
using the fine timing information of each hit photomultiplier 
(about 1 ns time resolution).   
An image obtained from a single telescope
gives only one angle: 
the projection of the direction on a plane normal to the mirror axis.
Stereo observations are required to obtain the two angles 
necessary to determine the direction of a gamma-ray in the sky.
For a single telescope,
the unknown angle of the shower direction lies on the extended 
long axis of the elliptic image.
However if we  know   both the the arrival times
and the height of the emitting point of \v Cerenkov light,
the shower direction  can be reconstructed.
In other words, the arrival timings
of \v Cerenkov photons emitted from a shower  can be calculated
if both the height and the direction of a shower are  known.
Therefore, by  assuming the height of shower generation 
($\sim 10,000 {\rm m}$ is used here),
the unknown  angle can be calculated using the observed
arrival timings of \v Cerenkov photons.
The simulation indicated that the obtained angle is  quite 
insensitive to the assumption of the shower height.
We actually obtained this angle event by event,
and used it only to estimate the direction of the shower development 
along the long axis of the image. 
By selecting the shower images for events from the target direction,
gamma-ray events should  be enhanced,
and indeed
$\alpha$ peaks in both the 1996 and 1997 data
became  more significant.
The result of the 1997 data is presented in Figs. \ref{fig:1}a and b
and the significance
of the $\alpha$ peaks\ was increased from 7.5 $\sigma$ to 8.8 $\sigma$ 
by this method.
After this cut, about 70\% of background events that survived through imaging
selections
can be further removed, while about 75\% of the gamma-ray events remained.

In order to obtain the differential energy spectrum,
several $\alpha$ plots were made 
by varying the minimum and maximum numbers of
detected \v Cerenkov  photons.
The collecting area, trigger efficiency and
threshold energy corresponding to each $\alpha$ plot
were independently estimated from simulations,
where  events were generated between 1 TeV and 30 TeV
using an initial power-law index of $\sim -2.2$.
This initial index was estimated from two observed integral fluxes 1996
(threshold energy $\ge$ 3.5 TeV) and 1997 (threshold energy $\ge$ 1.7 TeV).
At first the  spectrum was calculated using this index.
Using the new index obtained from this spectrum,
the above simulations and calculations  were  iterated in a similar way
until the spectral index converged.
Other initial values of the index were tried, 
but the resultant index converged to the similar value. 

\begin{figure}[t]
\vspace*{2.0mm} 
\includegraphics[width=8.3cm]{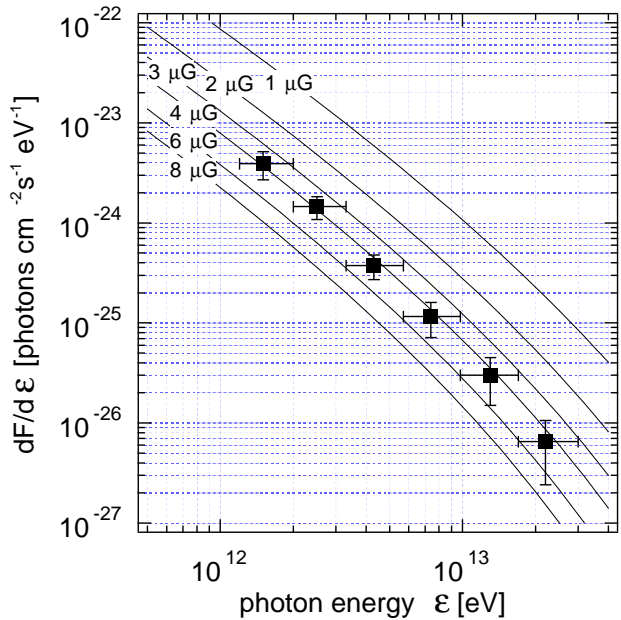} 
\caption{
Observed differential spectrum from the NE rim,
where only statistical errors are shown.
In addition, spectra of TeV gamma rays for several magnetic field
strengths {\it B} calculated by the IC model are presented.
}
\label{fig:2}
\end{figure}

The resultant differential spectrum, {\it J(E)}, between 1.5 TeV and 20 TeV
is plotted in Fig.\ref{fig:2} in which only statistical errors are shown.
It can be written as:

\begin{equation}
J(E)=(1.1\pm 0.4)
\times 10^{-11}(\frac{E}{1TeV})^{-2.3 \pm 0.2}
  \,\,{\rm TeV}^{-1}{\rm cm}^{-2}{\rm s}^{-1},
\end{equation}

\noindent where errors quoted are statistical.
Systematic errors are estimated to be $\sim$30\%.
Differential fluxes of 1996 and 1997 data were also individually calculated,
and those fluxes  were consistent within one sigma errors.
Furthermore the differential flux without applying the timing reconstruction
method was obtained to be quite similar to this flux.

\section{Discussion}

The resultant differential spectrum shows a spectral index of $-$2.3
between 1.5 TeV and 20 TeV.
The profile of the observed differential spectrum may  resolve
which IC or $\pi^{\circ}$ decay mechanism
dominates this TeV  emission.
The model used here originated
in Yoshida and Yanagita (1997), similar to the previous paper,
while
we improved the treatment of the approximation 
for the synchrotron flux (Naito et al., 1999):
a simple $\delta$ function centered on the characteristic
frequency was changed to the
full function describing synchrotron emission
(Rybicki and Lightman 1979 and references therein).
The electron spectrum was assumed to be as

\begin{equation}
\frac{dN_{e}}{dE}
= N_{0} {\left(\frac{E}{m_{e} c^2}\right)}^{s}
\exp \left(-\frac{E}{E_{\rm max}} \right) \ \ \ .
\end{equation}

In addition,
the latest data of both radio (Reynolds and Ellison 1992)
and X-rays from ROSAT and ASCA (Willingale et al.\ 1996; Ozaki 1998)
were used in fitting the synchrotron flux.
In particular, we used the partial fluxes emitted from the NE rim for all
data:
for the ASCA data the partial flux has been directly obtained by Ozaki
(1998),
and  for the radio and the ROSAT data the partial fluxes were estimated
using the morphological distributions of radio (Winkler and Long 1997)
and ROSAT (Willingale et al.\ 1996) data.
Comparing the calculated synchrotron flux to the observational data,
we obtain $s = -2.2$
and $(E_{\rm max}/TeV) \sqrt{(B/{\rm \mu G})} = 101$.
In Fig.\ref{fig:2} several expected IC spectra are plotted
by varying the strength of the ambient magnetic field,
where $B = 4 \pm 1 {\rm \mu G}$ is most probable,
and this model matches successfully in the convex shape.
The resultant field strength means the $E_{max} \sim 51$ TeV
from the above formula.

\begin{figure*}[t]
\vspace*{2.0mm} 
%\figbox*{}{}{\includegraphics*[width=12.0cm]{fig3.eps}}
\includegraphics[width=15.0cm]{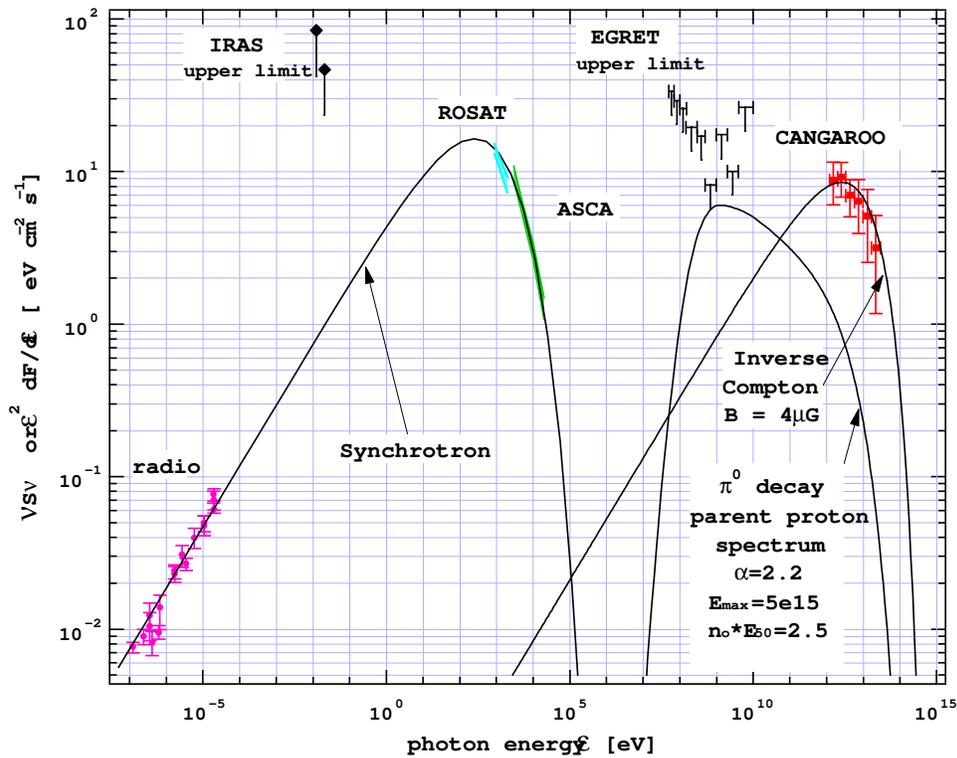} 
\caption{
Multi-band spectrum of energy fluxes observed
from the NE rim, where  observed fluxes or upper limits of radio (Reynolds
1998),
infrared, soft X-ray (estimated from Willingale et al. 1996),
hard X-ray (Ozaki 1998),
GeV gamma rays (calculated from the EGRET archive data),
and TeV gamma rays are presented.
Solid lines  are the fits based on the model of IC model
and neutral pion decay.
}
\label{fig:3}
\end{figure*}

Figure \ref{fig:3} shows the wide band energy spectrum from the radio to TeV regions
at the north rim of SN1006
and also the above fitting (Naito et al., 1999).
All data are fitted very well all over the wide band.
Here an allowable  spectrum due to  $\pi ^{\circ}$ decays generated by
high energy proton is also plotted
taking into account the upper limit of GeV gamma-ray fluxes,
which  obviously conflicts with the observed TeV gamma-ray flux.
Those upper limits in the GeV region of this figure were
calculated from the EGRET archive data using
a maximum likelihood method (Mattox et al., 1996).
These results seems very reasonable consequence considering  
the tenuous shell of ($\le \sim$0.4 cm$^{-3}$; Willingale et al. 1996).

Thus the identification of the parent particles of the TeV gamma-rays (electron
or proton) will be possible by observing the wide spectrum from sub- to 
multi-TeV region as shown in Fig.\ref{fig:3}.
A Gamma-ray spectrum flatter than $-$2.0 in this region is surely due to
the I.C. process, while that due to   $\pi^{\circ}$  decay generated by
collision between ISM and high energy proton is expected to be steeper
than $-$2.0.

From this fitting, the  energy flux of TeV gamma rays due to the IC process at 300 GeV 
is estimated to be $\sim$  4 eV cm$^{-2}$s$^{-1}$,
which is lower than the upper limit (8 eV cm$^{-2}$s$^{-1}$ at 300GeV)
reported in Chadwick et al.\ (2000). 
Also  we have observed  again TeV gamma rays from the NE rim of SN1006
observed by the 10m telescope as  presented in  this conference
(Hara et al., 2001).
Although the new data is still very preliminary,
the spectra are  consistent in the overlapping energy region. 

In this conference,
we present the interesting spectrum of TeV gamma rays
of PSR1706$-$44 showing an obvious break of the spectrum
around 1 TeV (Kushida et al., 2001).
This breaking can be  fit by the IC model quite well
similar to the case of SN1006.
These two results indicate the major role of IC process
for the production of high energy gamma rays in the universe.

However, our result showing the  IC process dominance for the production
of TeV gamma rays in SN1006 does  not imply  the nonexistence of 
the plenty of high energy protons accelerated by a diffusive shock.
Conventional diffusive shock theories naturally predict  
efficient acceleration of protons in SNRs.
In SN1006, a deficiency  of target protons surrounding the SNR
may conceal accelerated protons from being revealed.

RXJ1713.7$-$3946 
is the second SNR emitting both synchrotron X-rays and TeV
gamma rays.
However, the environment of RXJ1713.7$-$3946  is obviously different from 
SN1006; for example, association of RXJ1713.7$-$3946
with a molecular cloud is reported 
from the CO observation ( Slane et al., 1999).
Our new result on RXJ1713.7$-$3946  
indicates a more  intense flux below 1 TeV, 
while the flux above 1 TeV of it is similar to that of 
SN1006 (Enomoto et al., 2001).
Those features looks inconsistent with the scenario of the IC model.
Future  analyses of SN1006 and 
RXJ1713.7$-$3946 will soon provide
both the differential spectra between $\sim$ 300 GeV and $\sim$ 10 TeV
and the image of TeV gamma ray emission respectively, 
which will tell us what kind of  particle is accelerated and 
generates TeV gamma rays in each SNR.

\begin{acknowledgements}

\end{acknowledgements}
This work is supported by a Grant-in-Aid in Scientific Research
of the Japan Ministry of Education, Culture, Science, Sports 
and Technology, and
the Australian Research Council.

\end{document}